\documentclass[twocolumn]{aastex62}

\graphicspath{{./}{figures/}}

\received{February 28, 2024}
\revised{April 5, 2024}
\accepted{April 8, 2024}
\submitjournal{ApJL}

\shorttitle{Overmassive black holes at cosmic noon}
\shortauthors{Mezcua et al.}

\begin{document}

\title{Overmassive black holes at cosmic noon: linking the local and the high-redshift Universe}

\correspondingauthor{Mar Mezcua}
\email{marmezcua.astro@gmail.com}

\author[0000-0003-4440-259X]{Mar Mezcua}
\affiliation{Institute of Space Sciences (ICE, CSIC), Campus UAB, Carrer de Magrans, 08193 Barcelona, Spain}
\affiliation{Institut d'Estudis Espacials de Catalunya (IEEC), Edifici RDIT, Campus UPC, 08860 Castelldefels (Barcelona), Spain}

\author[0000-0001-9879-7780]{Fabio Pacucci}
\affiliation{Center for Astrophysics $\vert$ Harvard \& Smithsonian, Cambridge, MA 02138, USA}
\affiliation{Black Hole Initiative, Harvard University, Cambridge, MA 02138, USA}

\author[0000-0002-2536-1633]{Hyewon Suh}
\affiliation{Gemini Observatory/NSF's NOIRLab, 670 N. A'ohoku Place, Hilo, HI 96720, USA}

\author[0000-0002-2949-2155]{Malgorzata Siudek}
\affiliation{Institute of Space Sciences (ICE, CSIC), Campus UAB, Carrer de Magrans, 08193 Barcelona, Spain}
\affiliation{Instituto Astrofisica de Canarias, Av. Via Lactea s/n, E38205 La Laguna, Spain}

\author[0000-0002-5554-8896]{Priyamvada Natarajan}
\affiliation{Department of Astronomy, Yale University, New Haven, CT 06511, USA}
\affiliation{Department of Physics, Yale University, New Haven, CT 06520, USA}
\affiliation{Black Hole Initiative at Harvard University, 20 Garden St., Cambridge MA, 02138, USA}



\begin{abstract}
We report for the first time a sample of 12 supermassive black holes (SMBHs) hosted by low-mass galaxies at cosmic noon, i.e., in a redshift range consistent with the peak of star formation history: $z \sim 1-3$. These black holes are two orders of magnitude too massive for the stellar content of their hosts when compared with the local relation for active galaxies. These overmassive systems at cosmic noon share similar properties with the high-$z$ sources found ubiquitously in recent \textit{James Webb Space Telescope} (\textit{JWST}) surveys (same range of black hole-to-stellar mass ratio, bolometric luminosity, and Eddington ratio). We argue that black hole feedback processes, for which there is possible evidence in five of the sources, and the differing environments in galactic nuclei at these respective epochs play a key role in these overmassive systems. These findings contribute to our understanding of the growth and co-evolution of SMBHs and their host galaxies across cosmic time, offering a link between the early Universe ($z > 4$) observed by \textit{JWST} and observations of the present-day Universe ($z \lesssim 1$). 
\end{abstract}


 
\section{Introduction} 
\label{intro}
Supermassive black holes (SMBHs; $M_\mathrm{BH} > 10^{6}$ M$_{\odot}$) reside at the center of most massive (with typically a stellar mass $M_\mathrm{*} \sim10^{11}$ M$_{\odot}$) galaxies (e.g., \citealt{2013ARA&A..51..511K}) and grow through the accretion of matter and via coalescences during galaxy mergers (\citealt{2004ASSL..308..127N}; \citealt{2005LRR.....8....8M}; \citealt{2012Sci...337..544V}), which can ignite active accretion (i.e. active galactic nuclei, AGN). 

Tight correlations between BH mass and host galaxy properties (e.g., bulge mass, stellar velocity dispersion; e.g., \citealt{1998AJ....115.2285M}; \citealt{2000ApJ...539L...9F}; \citealt{2000ApJ...539L..13G}) strongly suggest that the growth of galaxies is linked to that of their central BHs (e.g., \citealt{1998A&A...331L...1S}), which can be accounted for if a fraction of the AGN energy output couples with the galactic medium and modulates star formation (thus regulating the growth of the host galaxy; see \citealt{2023NatAs...7.1376Z} for a recent review). Despite the key role SMBHs are now believed to play in galaxy formation and evolution, understanding their origin remains a challenge. The discovery of SMBHs as heavy as $10^{10}$ M$_{\odot}$ at redshifts $z\sim$ 6-7 ($\sim$700 Myr after the Big Bang; see \citealt{2023ARA&A..61..373F} for a review) and of $10^{6}$-$10^{7}$ M$_{\odot}$ at even earlier epochs ($z$ = 8.7, \citealt{2023ApJ...953L..29L}; $z$ = 10.3, \citealt{2024NatAs...8..126B}; $z$ = 10.6, 440 Myr after the Big Bang, \citealt{2024Natur.627...59M}) suggests these SMBHs could have started from seed BHs with masses in the range of $10^{2}$ - $10^{5}$ M$_{\odot}$ (also dubbed intermediate-mass BHs) as early as $z \sim$ 20 (e.g., \citealt{2010A&ARv..18..279V}; \citealt{2017IJMPD..2630021M}; \citealt{2020ARA&A..58..257G}; \citealt{2022MNRAS.509.1885P}). Broadly speaking, two classes of seeding models - light and heavy - have been proposed and studied in detail (e.g., \citealt{2014GReGr..46.1702N}). Light seed BHs of $\sim$100 M$_{\odot}$ are expected to have formed from the  death of the first generation of Population III stars; while heavier seed BHs of $\sim10^{4}-10^{5}$ M$_{\odot}$ could have formed via direct collapse of pre-galactic gas (e.g., \citealt{1994ApJ...432...52L};  \citealt{2006MNRAS.371.1813L}). Other possibilities include early super-Eddington growth (\citealt{2005ApJ...633..624V}), stellar mergers in early nuclear stellar clusters (e.g., \citealt{1999A&A...348..117P}; \citealt{2009ApJ...694..302D}), rapid growth of light seeds via wind-fed accretion in early nuclear star clusters (\citealt{2014Sci...345.1330A}), and primordial BHs (e.g., \citealt{2022ApJ...926..205C}; \citealt{2022arXiv221213903Z}). There is growing evidence that the recent detection of $10^{6}$-$10^{7}$ M$_{\odot}$ BHs at $z >$ 8 favors the heavy seeding scenario, as lighter seeds would require periods of super-Eddington accretion (e.g., \citealt{2022MNRAS.509.1885P}; \citealt{2024NatAs...8..126B}; \citealt{2023ApJ...953L..29L}; \citealt{2024Natur.627...59M}; \citealt{2024ApJ...960L...1N}).

At low redshifts, insights into seeding can be obtained from a multiplicity of observational probes, ranging from the low-mass end of the BH occupation fraction (e.g., \citealt{2018MNRAS.481.3278R}; \citealt{2023MNRAS.518.1880B}; \citealt{2023MNRAS.523.5610B}) to the BH mass function (e.g., \citealt{2019ApJ...883L..18G}; \citealt{2021MNRAS.503.1940H}; \citealt{2022ApJ...924...56S}) and luminosity function (e.g, \citealt{2008MNRAS.383.1079V}; \citealt{2023MNRAS.523.5610B}). These studies triggered a quest for actively accreting intermediate-mass BHs in dwarf galaxies, where they can be detected in large numbers as low-mass ($M_\mathrm{BH} \lesssim 10^{6}$ M$_{\odot}$) AGN (e.g., see review by \citealt{2022NatAs...6...26R}). The use of optical spectroscopy has thus far yielded detections of hundreds of AGN in dwarf galaxies and provided estimates of the AGN occupation fractions (taken as a proxy for BH occupation fraction) ranging from $< 1$ \% (\citealt{2013ApJ...775..116R}; \citealt{2022ApJ...937....7S}) to $\sim 20$ \% (\citealt{2022ApJ...931...44P}; \citealt{2020ApJ...898L..30M,2024MNRAS.tmp..271M}). AGN variability studies of large samples of nearby sources are also providing interesting constraints on the local BH occupation fractions (\citealt{2024arXiv240206882B}). AGN fractions of $<$1\% corrected for completeness have been derived from X-ray surveys (e.g., \citealt{2018MNRAS.478.2576M}; \citealt{2020MNRAS.492.2268B,2022MNRAS.510.4556B}; \citealt{2024MNRAS.527.1962B}). Meanwhile, theoretical predictions for the AGN fraction in dwarf galaxies have also been derived (e.g, \citealt{2021ApJ...920..134P}). While these estimated AGN fractions seem to favor heavy seeding BH models, a direct link between the early seeds and the BHs powering AGN in low-redshift dwarf galaxies remains to be firmly established.


Recently, \cite{2023MNRAS.518..724S} identified a sample of 4,315 AGN in dwarf galaxies at 0.5 $< z <$ 0.9, the largest of its kind. The average BH mass of this sample is log $M_\mathrm{BH}$ = 8.2 M$_{\odot}$, derived from  correlations between narrow emission lines (\citealt{2019MNRAS.487.3404B}). This average BH mass is two orders of magnitude more massive than expected from local BH-galaxy scaling relations. Based on broad emission lines, \cite{2023ApJ...943L...5M} recently identified another sample of overmassive BHs in dwarf galaxies at $z \sim$0.4-0.9. Interestingly, the \textit{James Webb Space Telescope} (\textit{JWST}) in its first year of operation has found tens of such overmassive BHs at $z = 4-10$, which lie $10-100$ times above the local $M_\mathrm{BH}$-$M_\mathrm{*}$ relation (\citealt{2023ApJ...957L...3P}). Many of these overmassive SMBHs are hosted in low-mass galaxies of $M_\mathrm{*} \sim 10^8-10^{10} \, M_\odot$ (e.g., \citealt{2024NatAs...8..126B}; \citealt{2024Natur.628...57F}; \citealt{2023ApJ...959...39H}; \citealt{2023ApJ...954L...4K}; \citealt{2023ApJ...957L...7K}; \citealt{2023arXiv230801230M,2024Natur.627...59M}; \citealt{2023ApJ...953..180S}; \citealt{2023A&A...677A.145U}). 

The finding that these BHs are overmassive with respect to the local $M_\mathrm{BH}$-$M_\mathrm{*}$ relation is not a consequence of survey selection effects, but is rather expected from some theoretical simulations (e.g., \citealt{2013MNRAS.432.3438A}; \citealt{2022ApJ...927..237I}; \citealt{2022MNRAS.511..616T}), especially if the SMBHs arise from heavy seeds (e.g., \citealt{2023MNRAS.519.2155S}). Recent results from the ASTRID, Illustris TNG50 simulation suites also predict the existence of such an overmassive SMBH population in the local Universe (\citealt{2023MNRAS.522.4963W}). At these late epochs, environmental effects like tidal stripping of the stellar component of host galaxies are implicated in causing the central SMBHs in dwarf galaxies to be overmassive compared to observed scaling relations (e.g., \citealt{2018MNRAS.473.1819F,2021MNRAS.506.4702F}).

In this Letter we report on a sample of 12 low-mass galaxies hosting overmassive BHs at $z \sim 1-3$, constituting the first such sample at the peak of cosmic star formation history (referred to as the `cosmic noon'; \citealt{2014ARA&A..52..415M}). These $z \sim 1-3$ overmassive BHs and the $z \sim 0.4-0.9$ from \cite{2023ApJ...943L...5M} share the same properties as the $z > 4$ \textit{JWST} overmassive BHs, allowing us to study for the first time low-mass galaxies in the high-redshift Universe with lower redshift counterparts and to probe BH-galaxy co-evolution across cosmic time.
The sample and data analysis are described in Sect.~\ref{sample}. The results obtained are reported in Sect.~\ref{results}. Discussion and conclusions are provided in Sect.~\ref{conclusions}. We adopt a $\Lambda$-CDM cosmology with $H_{0}=70$ km s$^{-1}$ Mpc$^{-1}$, $\Omega_{\Lambda}=0.73$ and $\Omega_{m}=0.27$.

\section{Sample and Analysis}
 \label{sample}
The sample selection is performed following the same procedure as in \cite{2023ApJ...943L...5M}, which we summarize here.

\subsection{Stellar mass measurements}
We start from a parent sample of 1,161 broad-line AGN galaxies identified in the VIMOS Public Extragalactic Redshift Survey (VIPERS), which includes $\sim$90,000 spectra ranging from $z$ = 0.1 to $z$ = 4.56 (\citealt{2018A&A...609A..84S}). We fit the multiwavelength spectral energy distribution (SED) of these 1,161 galaxies from ultraviolet to infrared wavelengths using a modified version of the SED code of  \cite{2019ApJ...872..168S}. This includes a combination of galaxy and AGN templates, using the same SED libraries as in \textsc{AGNfitter} \citep{2016ApJ...833...98C}. The best fit is initially determined using a $\chi^2$ minimization. We note that the delineation of the flux contribution from the stellar component versus the AGN is extremely challenging and here we do it adopting multiple independent empirical methods.

For those sources with an X-ray detection in the XMM-XXL catalogue (\citealt{2018A&A...620A..12C}; 452 out of the 1,161 broad-line AGN), we confirm that the rest-frame luminosity at 2500\AA\ of the best-fitting AGN component correlates with the X-ray luminosity, in agreement with the X-ray-to-ultraviolet correlation of AGN (e.g., \citealt{2016ApJ...819..154L}). 

To account for the degeneracies inherent to the SED fitting, we derive a probability distribution function (PDF) for the stellar mass that considers an AGN fraction (f$_\mathrm{AGN}$) ranging from 0 to 1. We then perform the SED fitting assuming that the galaxy emission dominates over the AGN in the K-band, using the oldest stellar population possible in order to obtain a conservative upper limit on the best-fit stellar mass. The difference between the highest most probable value (MsPDF) and the best-fit stellar mass (MsBEST) is then taken as the uncertainty in the stellar mass (see Fig.~\ref{SEDfit}, top right panel). An additional 0.2 dex is added to the uncertainties in order to account for differences in the stellar population models arising from factors such as the choice of the Initial Mass Function. The star formation rate (SFR) is derived from the best-fit SED as SFR $\propto e^{t/\tau}$, where the characteristic time of the exponentially decaying star formation histories of the stellar population models range from $\tau$=0.1 to 30 Gyr.

Since SED-fitted stellar masses are typically highly dependent on the SED fitting code, we also derive the stellar masses independently using the Code Investigating GALaxy Emission (CIGALE; \citealt{2019A&A...622A.103B}) and adopting different parameters for the models of the stellar populations and star formation history, dust emission and attenuation, and AGN emission. For those sources with X-ray emission, we also perform the SED fitting with the X-CIGALE code (\citealt{2020MNRAS.491..740Y}), which incorporates X-ray fluxes. We find that even with these different parameterizations the stellar masses are consistent with those derived from \textsc{AGNfitter}. More details on the SED fitting can be found in the Appendices A and B in \cite{2023ApJ...943L...5M}.

The sample of low-mass galaxies reported here are selected as having $z > 1$ (those sources with $z < 1$ are reported in  \citealt{2023ApJ...943L...5M}) and an MsBEST of log $M_\mathrm{*} \leq$ 9.5 M$_{\odot}$, which is the typical threshold considered in studies of AGN in dwarf galaxies. This mass range corresponds approximately to the stellar mass of the Large Magellanic Cloud (e.g., \citealt{2013ApJ...775..116R,2020ApJ...888...36R}; \citealt{2016ApJ...817...20M,2018MNRAS.478.2576M,2019MNRAS.488..685M}; \citealt{2020ApJ...898L..30M,2024MNRAS.tmp..271M}). We consider only those galaxies with an upper MsPDF not extending beyond 10$^{10}$ M$_{\odot}$ to restrict our sample to the low-mass regime. We note that, because of the large uncertainties in the stellar masses at intermediate to high redshifts, it is more appropriate to use the term `low-mass' rather than `dwarf' to refer to those galaxies with $M_\mathrm{*} < 10^{10}$ M$_{\odot}$ found at $z \gtrsim 1$, as at these redshifts we can only confidently distinguish between low-mass and massive galaxies. This nomenclature we argue applies not only to this Letter but also to those samples of low-mass galaxies derived using \textit{JWST} data. From now on, the term `dwarf' will therefore be used to refer only to local sources.

\subsection{Emission line fitting}
The cuts in redshift and stellar mass applied above yield a sample of 13 low-mass AGN galaxies at $z > 1$. To confirm the presence of broad emission lines in these sources, we use the public Python QSO fitting code (PyQSOFit; \citealt{2018ascl.soft09008G}) to fit their optical spectrum. PyQSOFit fits the continuum emission with a power-law using a few emission-line-free regions and then subtracting it. The rest-frame MgII and CIV emission lines are then fitted using typically two components, one broad and one narrow, where the narrow one is defined as having a Full Width at Half Maximum (FWHM) $<$ 1,200 km s$^{-1}$ (\citealt{2019ApJS..241...34S}). As a result, PyQSOFit outputs the line flux, FWHM, equivalent width and dispersion of the broad and narrow components as well as the continuum luminosity at $1350 \rm \AA$ and $3000 \rm \AA$ (when available). A Monte Carlo approach is used to compute the uncertainties in the emission line measurements. 

For five of the 13 low-mass AGN galaxies, SDSS\footnote{Sloan Digital Sky Survey, \url{https://www.sdss.org}} DR14 spectra are also available. We fit these SDSS spectra using the same procedure as described above. We favor the SDSS fit over that of VIPERS for three sources, for which the VIPERS spectrum contains only one emission line (e.g., only the CIV line is in the VIPERS spectrum while in the SDSS there are additional lines, namely, CIV and Ly$\alpha$) and the $\chi^2$ of the SDSS fit is better than the one derived from VIPERS. For one of these three sources, 401126746, the CIV line of the SDSS spectrum shows an artifact that prevents a reliabe use of the FWHM to derive the BH mass. For this source, the BH mass is derived from the FWHM and  luminosity of the H$\alpha$ emission line (\citealt{2020ApJ...889...32S}) detected in Gemini/GNIRS Fast Turnaround observations (PI: Suh; see Appendix \ref{Gemini}).

Five of the sources show possible asymmetric components in the Ly$\alpha$, CIV or CIII] emission suggestive of outflows (see e.g., Fig.~\ref{SEDfit}). A thorough study of AGN outflows is out of the scope of this paper. However, in order to avoid any biases in the BH mass estimations, for these galaxies the fit of the CIV emission line is attempted with one additional broad component in order to take the possible outflow into account. For four out of the five sources, this does not improve the fit. For the remaining source, 127008752, the addition of a broad component yields a slight improvement of the fit ($\chi^2$ = 0.9 versus $\chi^2$ = 1.4 when only using one broad component) but also increases the FWHM of the broad component used to derive the BH mass by a factor 1.2. To be as conservative as possible, we thus proceed with the fit that provides the lowest value of FWHM. 

In total, we find that the optical VIPERS, SDSS, or near-infrared Gemini/GNIRS spectrum are reliably fitted by PyQSOFit for 12 out of the 13 low-mass AGN galaxies. Our final sample is thus composed of 12 AGN low-mass galaxies at $z>$1. The spectral fit of one of these 12 sources is shown in the Appendix, Fig.~\ref{SEDfit}, middle panel.

\subsection{Black hole masses}
BH masses are derived from the width of the MgII (if $z < 2$) or CIV or H$\alpha$ (if $z > 2$) broad emission line components and the line luminosity or adjacent continuum at 1350 \AA\ or 3000 \AA, when available. We use the single-epoch virial calibrations from \cite{2006ApJ...641..689V} and \cite{2012ApJ...753..125S} based on a mean virial factor $\epsilon \sim$1 (e.g., \citealt{2004ApJ...615..645O}; \citealt{2013ApJ...773...90G}) and with a scatter of $\sim$0.3 dex (e.g., \citealt{2012ApJ...753..125S}). Adding in quadrature the measurement uncertainties of $\sim$0.1 dex results in a total BH mass uncertainty of $\sim$0.4 dex. A word of caution is warranted regarding this methodology for estimating the BH mass for high-$z$ objects. Single-epoch virial calibrations, such as the one from \cite{2006ApJ...641..689V}, are calibrated in the local Universe, typically at $z \ll 1$. Hence, while this constitutes the only viable methodology available with current data, additional systematic effects could be present in the BH mass estimates. This should however not threaten the results here reported, which for most sources should hold unless the BH masses are overestimated by a factor $\sim 60$ (\citealt{2023ApJ...957L...3P}).

AGN bolometric luminosities are derived from the continuum luminosities at $1350 \rm \AA$ and $3000 \rm \AA$ applying the bolometric correction factors of \cite{2006ApJS..166..470R}. Eddington rates are then derived using $\lambda_\mathrm{Edd}$ = $L_\mathrm{bol}$/($M_\mathrm{BH} \times 1.3 \times 10^{38}$).

\section{Results}
\label{results}
The new sample of 12 low-mass AGN galaxies here reported have spectroscopic redshifts ranging from $z$ = 1.32 to 2.78, constituting the first such sample at the key epoch of cosmic noon ($z \sim$ 1-3) where cosmic star formation activity and BH growth reached their peaks (e.g., \citealt{2014ARA&A..52..415M}). The SFRs of this `cosmic noon' sample of 12 low-mass galaxies are in the range log SFR = 1.1 to 1.7  M$_{\odot}$ yr$^{-1}$, which locates them above the main-sequence of star-forming galaxies (e.g., \citealt{2014ApJ...795..104W}). Five of the sources have a 0.2-12 keV X-ray counterpart in the 4XMM-DR13 catalogue (\citealt{2020A&A...641A.136W}), with a k-corrected luminosity ranging log L$_\mathrm{0.2-12 keV}$ = 44.2 - 44.8 erg s$^{-1}$ that confirms their AGN nature. We note that, given the estimated bolometric luminosity of these sources (i.e., log L$_\mathrm{bol} \sim$ 45-46 erg s$^{-1}$, see Table~\ref{table1}), this leads to an average hard X-ray bolometric correction of $k_\mathrm{X} \sim 20$, which is in perfect agreement with the universal bolometric corrections derived in e.g., \cite{2020A&A...636A..73D}. The most relevant information about each low-mass AGN galaxy is provided in Table~\ref{table1}.

\begin{table*}
\begin{center}
\caption{Properties of the `cosmic noon' sample of 12 low-mass galaxies at $z \sim 1-3$ hosting overmassive BHs.}
\label{table1}
\begin{tabular}{cccccccccc}
\hline
\hline 
VIPERS &  R.A.          & Dec.        & $z$     &  log $M_\mathrm{*}$ & log $M_\mathrm{BH}$  & log $L_\mathrm{bol}$  & Survey & Broad& log $L_\mathrm{X}$  \\
ID 	     & (J2000)	     &  (J2000)   &	        & (M$_{\odot}$)       & (M$_{\odot}$)        & (erg s$^{-1}$)        &        & Line & (erg s$^{-1}$) \\
(1) & (2)   & (3)   & (4)   & (5)   & (6)   & (7)   &  (8) & (9) & (10)  \\
\hline
402088938 & 22:07:12.95 & +01:11:15.6 &   1.322 & 9.1$^{+0.5}_{-0.1}$ & 8.6$\pm$0.3 &    45.0 &   VIPERS &      MgII &           --       \\
103212570 & 02:10:19.15 & -05:38:55.4 &   1.384 & 9.4$^{+0.5}_{-0.2}$ & 8.9$\pm$0.3 &    45.2 &   VIPERS &      MgII &           --       \\
116100529 & 02:26:07.33 & -05:06:44.2 &   1.628 & 9.4$^{+0.9}_{-0.1}$ & 7.3$\pm$1.2 &    45.2 &   VIPERS &      MgII &           44.2$\pm$0.3       \\
117162807 & 02:28:22.40 & -04:49:59.1 &   1.628 & 9.0$^{+0.7}_{-0.0}$ & 8.4$\pm$0.3 &    45.4 &   VIPERS &      MgII &           44.5$\pm$1.6       \\
113197834 & 02:13:08.20 & -04:43:06.5 &   2.565 & 9.5$^{+0.8}_{-0.1}$ & 8.1$\pm$0.4 &    46.1 &     SDSS &       CIV & 44.7$\pm$0.3       \\
401126746 & 22:02:56.52 & +01:21:56.5 &   2.587 & 9.2$^{+0.7}_{-0.1}$ & 8.8$\pm$0.3 &    45.3 &   SDSS/Gemini &  CIV/H$\alpha$ &           --       \\
405118143 & 22:17:53.28 & +01:22:02.8 &   2.614 & 9.5$^{+0.6}_{-0.1}$ & 7.7$\pm$0.3 &      -- &   VIPERS &       CIV &           --       \\
126027215 & 02:28:46.40 & -04:32:34.6 &   2.696 & 9.2$^{+1.0}_{-0.2}$ & 7.9$\pm$0.3 &      -- &   VIPERS &       CIV &           44.8$\pm$0.2       \\
112102440 & 02:10:15.93 & -05:08:32.8 &   2.738 & 9.0$^{+1.1}_{-0.2}$ & 8.4$\pm$0.3 &      -- &   VIPERS &       CIV &           44.7$\pm$0.2       \\
411013198 & 22:15:28.50 & +01:48:15.1 &   2.744 & 9.5$^{+0.6}_{-0.1}$ & 8.3$\pm$0.4 &    45.9 &     SDSS &       CIV &           --       \\
111011166 & 02:07:37.71 & -05:34:31.7 &   2.744 & 9.4$^{+0.8}_{-0.1}$ & 8.0$\pm$0.3 &      -- &   VIPERS &       CIV &           --       \\
127008752 & 02:32:35.83 & -04:39:16.2 &   2.768 & 9.2$^{+0.8}_{-0.1}$ & 8.3$\pm$0.3 &      -- &   VIPERS &       CIV &           --       \\
\hline
\hline
\end{tabular}
\end{center}
\smallskip\small {\bf Column designation:}~(1) VIPERS ID; (2,3) right ascension and declination; (4) redshift; (5) stellar mass derived from SED fitting; (6) BH mass derived from single-epoch virial calibrations; (7) bolometric luminosity derived from the monochromatic continuum luminosity at 1350 \AA\ or 3000 \AA; (8) survey for spectroscopy; (9) fitted broad emission line; (10) 0.2-12 keV X-ray luminosity, if available. The uncertainties in the stellar mass include a 0.2 dex to account for differences in the stellar population models. The uncertainties in BH mass are the quadratic sum of the measurement uncertainties ($\sim$0.1 dex) and the systematic uncertainties carried by single-epoch virial calibrations ($\sim$0.3 dex).
\end{table*}

\subsection{AGN properties}
The `cosmic noon' sample of 12 low-mass AGN galaxies at $z \sim 1-3$ have log $M_\mathrm{BH} = 7.3-8.9$ M$_{\odot}$ with an average uncertainty of 0.4 dex. These SMBH masses are two to three orders of magnitude higher than those of the low-mass AGN found in local dwarf galaxies ($M_\mathrm{BH} \lesssim 10^{6}$ M$_{\odot}$; e.g., \citealt{2013ApJ...775..116R}; \citealt{2020ApJ...898L..30M,2024MNRAS.tmp..271M}). The bolometric luminosities of the `cosmic noon' sample range from log L$_\mathrm{bol}$ = 44.9 to 46.1 erg s$^{-1}$ and are again orders of magnitude higher than those of AGN dwarf galaxies in the local Universe (log L$_\mathrm{bol} \sim$ 40-42 erg s$^{-1}$; e.g., \citealt{2020ApJ...898L..30M,2024MNRAS.tmp..271M}). The Eddington ratios are in the range $\lambda_\mathrm{Edd}$ = 0.02 - 0.8, with a median value $\lambda_\mathrm{Edd}$ = 0.2, indicating that the AGN low-mass galaxies at $z \sim 1-3$ are mostly accreting at sub-Eddington rates.

The BH masses, bolometric luminosities, and Eddington ratios of the low-mass AGN galaxies at $z \sim 1-3$ are very similar to those of the seven AGN dwarf galaxies found by \cite{2023ApJ...943L...5M} at $z = 0.35-0.93$ (log $M_\mathrm{BH}$ = 7.6-8.7 M$_{\odot}$, log L$_\mathrm{bol}$ = 44.8 to 45.4 erg s$^{-1}$, median $\lambda_\mathrm{Edd}$ = 0.1). The galaxy properties of both samples are also very similar (same range of stellar masses and of SFR), with all the sources being star-forming galaxies (see \citealt{2023ApJ...943L...5M}). All together indicates that the `cosmic noon' AGN low-mass galaxies and those at $z < 1$ are similar sources just observed at different cosmic epochs. 

\subsection{The $M_\mathrm{BH}$-$M_\mathrm{*}$ scaling relation}
\begin{figure*}
\centering
\includegraphics[width=\textwidth]{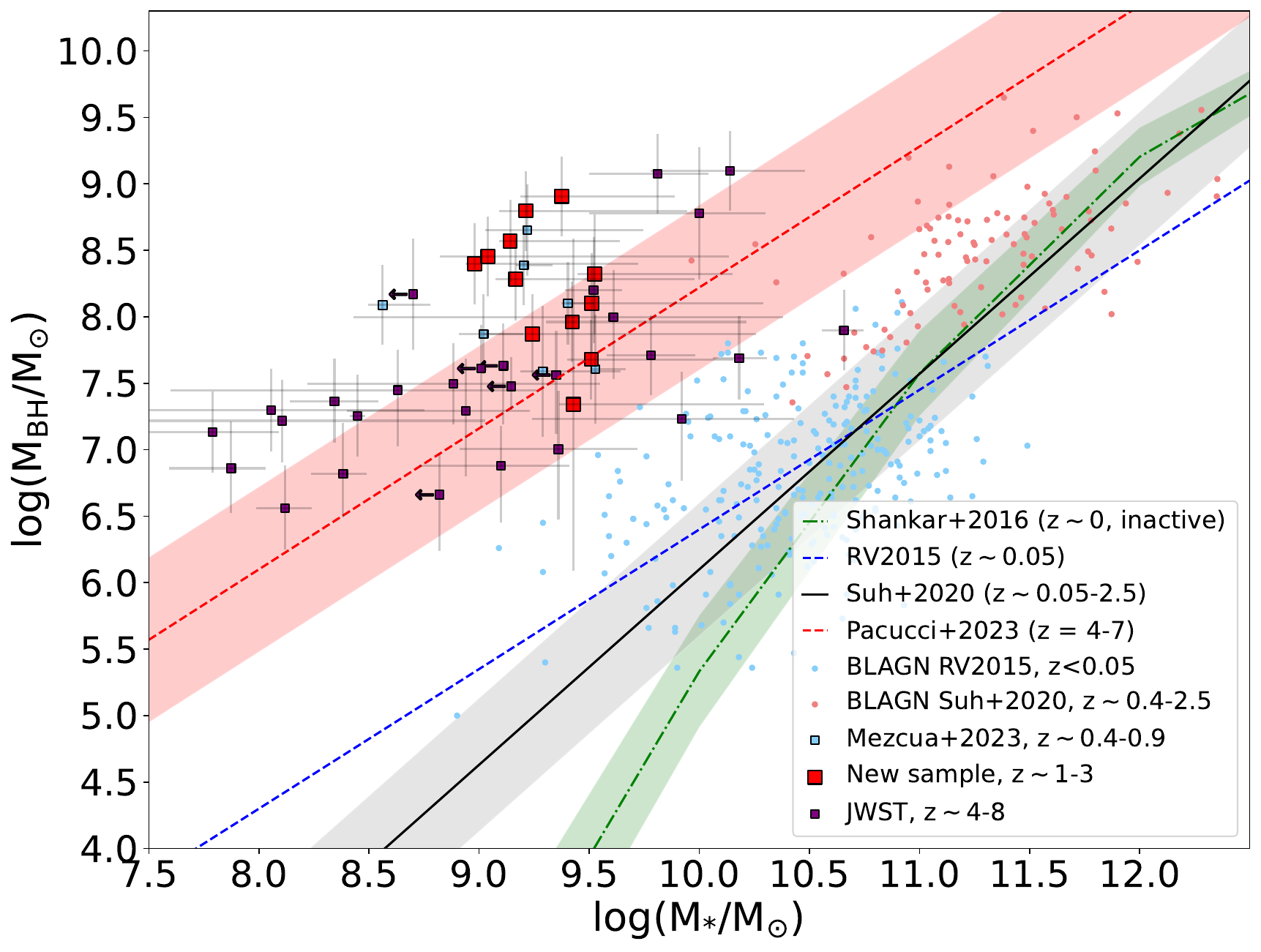}
\caption{$M_\mathrm{BH}$ versus $M_\mathrm{*}$ for the VIPERS sample of AGN dwarf galaxies at $z<1$ of \cite{2023ApJ...943L...5M} (blue squares), the new `cosmic noon' sample of AGN in low-mass galaxies here reported (red squares), and a compilation of \textit{JWST} AGN in low-mass galaxies at $z>4$ (purple squares). We also show for comparison the low-z AGN of \cite{2015ApJ...813...82R} (RV2015, blue dots) and the intermedidate-z AGN of \cite{2020ApJ...889...32S} (Suh+2020, red dots), whose masses have been computed using the same procedure and parameters as the VIPERS samples. The solid black line shows the local + intermediate-z $M_\mathrm{BH}$-$M_\mathrm{*}$ correlation found for the combination of the RV2015 and Suh+2020 samples with a 1$\sigma$ scatter of 0.5 dex. The red dashed line shows the high-z $M_\mathrm{BH}$-$M_\mathrm{*}$ correlation found by \cite{2023ApJ...957L...3P} for AGN found at $z>4$ by \textit{JWST} based on the detection of broad H$\alpha$ or H$\beta$ emission. The green dashed-dotted line shows the \cite{2016MNRAS.460.3119S} local correlation for inactive galaxies.}
\label{MbhMstellar}
\end{figure*}

The AGN dwarf galaxies at $z \sim 0.4-0.9$ of \cite{2023ApJ...943L...5M} were found to host BHs more massive than expected from the local $M_\mathrm{BH}$-$M_\mathrm{*}$ scaling relation of AGN (e.g., at $z < 0.05$, \citealt{2015ApJ...813...82R}; from now on RV2015) and that at $z \sim 0.4-2.5$ of \cite{2020ApJ...889...32S} (from now on Suh+2020). To investigate further the `cosmic noon' sample of 12 AGN low-mass galaxies at $z \sim 1-3$ here reported, we locate them in the $M_\mathrm{BH}$-$M_\mathrm{*}$ diagram and compare them to the RV2015 sample of AGN in dwarf and massive galaxies, the Suh+2020 sample of AGN in massive galaxies, and the \cite{2016MNRAS.460.3119S} local correlation for inactive galaxies corrected for resolution-related effects (see Fig.~\ref{MbhMstellar}). As for the \cite{2023ApJ...943L...5M} sources, the new sample at $z \sim 1-3$ is also found to be overmassive with respect to the stellar mass according to the local scaling relation for AGN. The BH mass offset ($\Delta M_\mathrm{BH}$) from the $M_\mathrm{BH}$-$M_\mathrm{*}$ can be derived using a Monte Carlo approach, by assigning 100 random variables to the $M_\mathrm{BH}$ and $M_\mathrm{*}$ distributions of each source and calculating $\Delta M_\mathrm{BH}$ based on the distribution of 100$^2$ possibilities over the number of sources (e.g., \citealt{2018MNRAS.474.1342M}; see Appendix D in \citealt{2023ApJ...943L...5M}). For the seven overmassive BHs in \cite{2023ApJ...943L...5M} at $z = 0.35-0.93$, the median of the BH mass offset from the local + intermediate-z $M_\mathrm{BH}$-$M_\mathrm{*}$ correlation of Suh+2020 was $\Delta M_\mathrm{BH}$ = 3.2 $\pm$ 1.3 with a significance of 100\% (3$\sigma$ level). The same offset and significance are also now obtained for the new sample of 12 AGN in low-mass galaxies at $z \sim 1-3$, indicating that these sources at cosmic noon are also overmassive. Combining the seven $z<1$ sources of \cite{2023ApJ...943L...5M} with the new `cosmic noon' sample of 12 sources, we also find the same $\Delta M_\mathrm{BH}$ but at a 5$\sigma$ level, suggesting once more that the in total 19 AGN in low-mass galaxies found from $z = 0.35$ to $z = 2.7$ in the VIPERS survey share the same properties. When considering the local relation of inactive galaxies corrected for resolution-related selection effects of \cite{2016MNRAS.460.3119S}, the sources are still as offset as they are from the local scaling relation for AGN; this is a result that will be further discussed in Sect.~\ref{conclusions}.


\subsection{Overmassive BHs across cosmic time}
\begin{figure}
\centering
\includegraphics[width=0.47\textwidth]{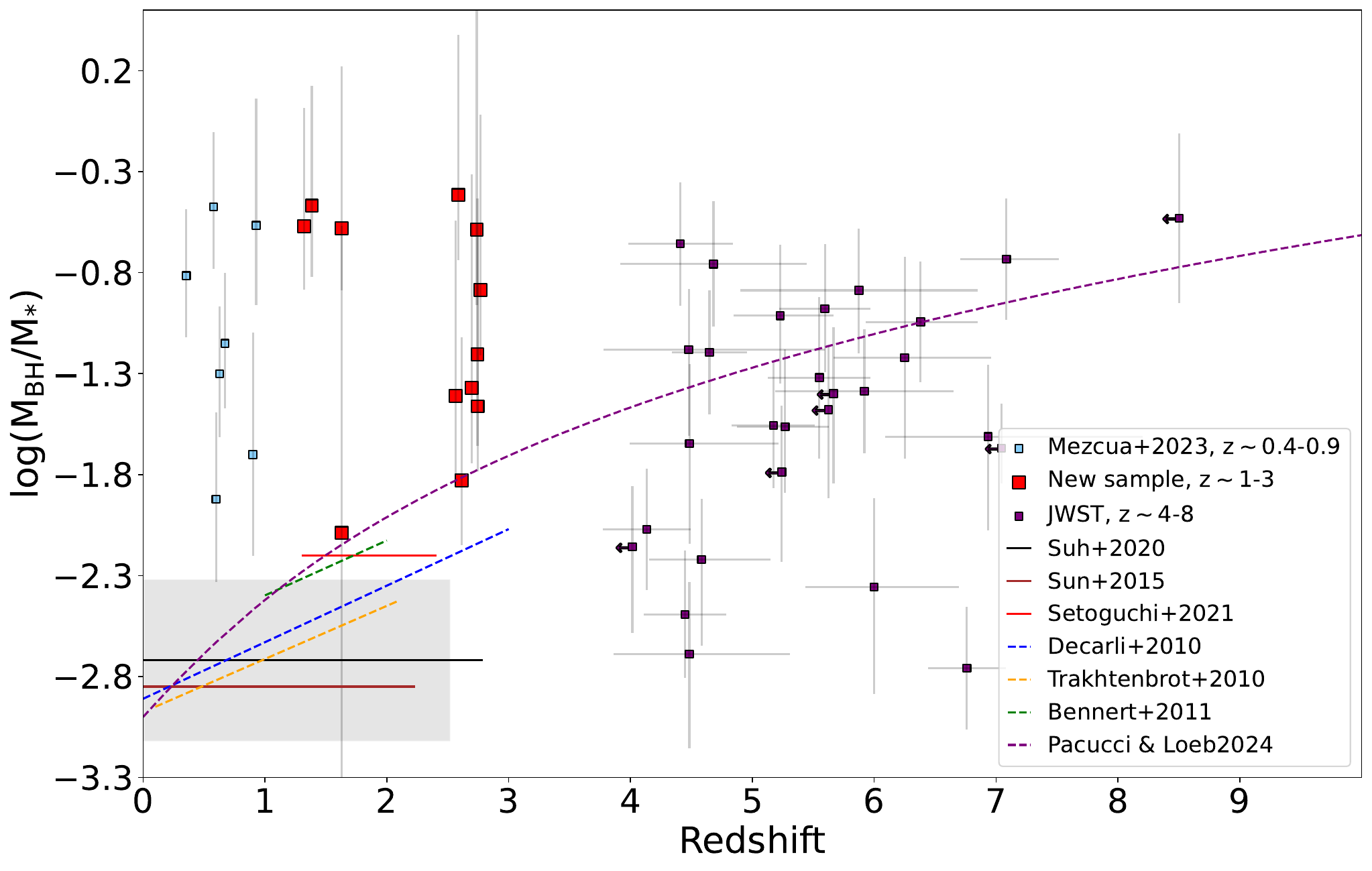}
\caption{Ratio of $M_\mathrm{BH}$/$M_\mathrm{*}$ versus redshift for the VIPERS sample of AGN dwarf galaxies at $z<1$ of \cite{2023ApJ...943L...5M}, the new VIPERS `cosmic noon' sample of AGN in low-mass galaxies here reported, and a compilation of \textit{JWST} AGN in low-mass galaxies at $z > 4$. We show for comparison the non-evolution found by \cite{2020ApJ...889...32S} for $z \sim 0-2.5$ (black line, 1$\sigma$ scatter$\sim$0.5 dex), \cite{2015ApJ...802...14S} for $z \sim 0-2$ (brown line), and \cite{2021ApJ...909..188S} for $z \sim 1.2-1.7$ (red line), and the z-evolution found by \cite{2010MNRAS.402.2453D} for $z \sim 0-3$ (dashed blue line), \cite{2010MNRAS.406L..35T} for $z \sim 0.1-2$  (dashed yellow line), \cite{2011ApJ...742..107B} for $z \sim 1-2$ (dashed green line, including the data of \citealt{2010ApJ...708..137M}), and \cite{2024ApJ...964..154P} for the \textit{JWST} $z > 4$ population of overmassive BHs (dashed purple line). We note that the Suh+2020 sample is the only one for which the BH and stellar masses have been computed using the same procedure and parameters as the VIPERS samples.}
\label{MbhMstellarevolution}
\end{figure}

\begin{figure}
\centering
\includegraphics[width=0.47\textwidth]{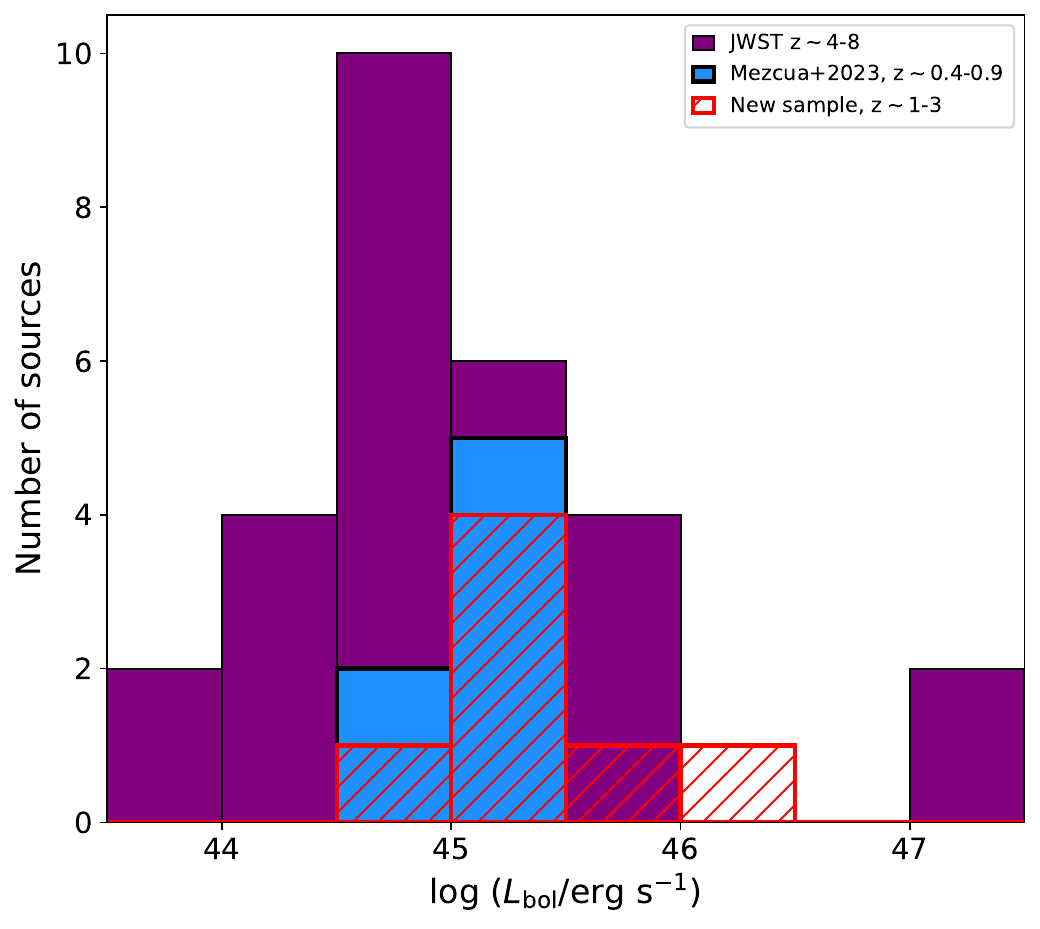}
\includegraphics[width=0.47\textwidth]{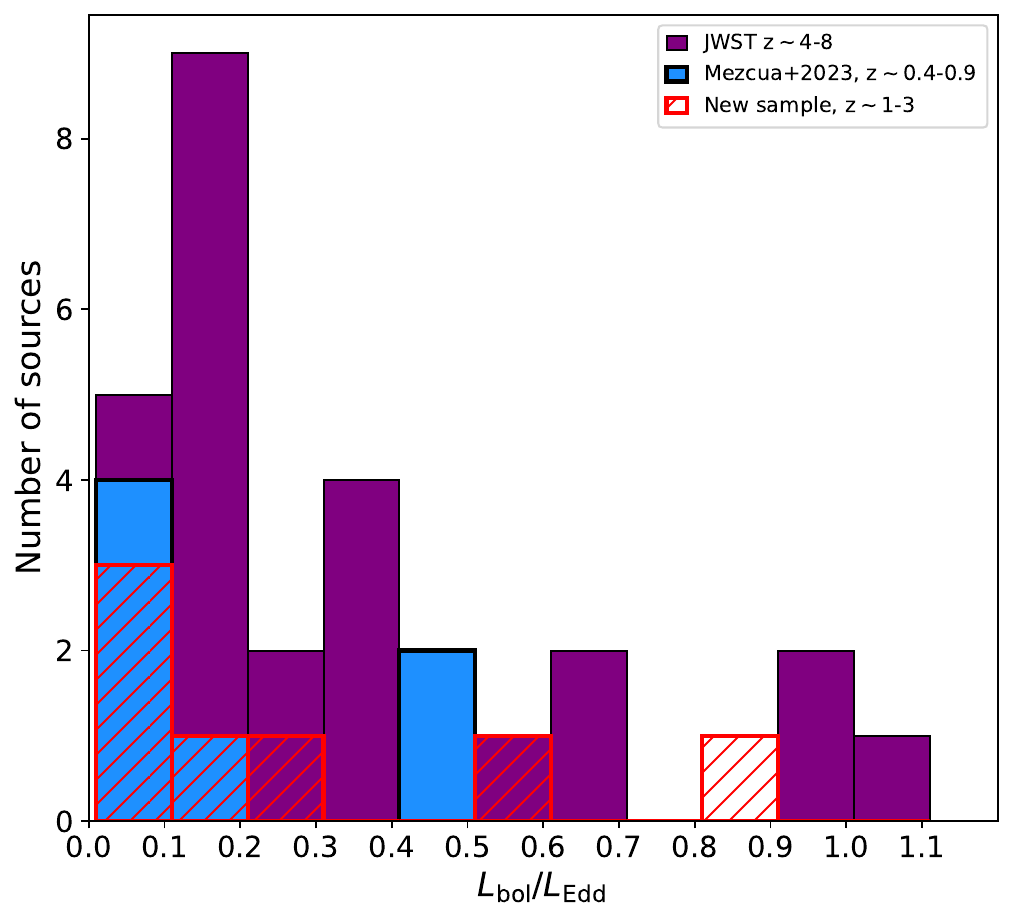}
\caption{Distribution of bolometric luminosity (top) and Eddington ratio (bottom) for the sample of AGN dwarf galaxies at $z < 1$ of \cite{2023ApJ...943L...5M}, the new `cosmic noon' sample of AGN in low-mass galaxies here reported, and a compilation of \textit{JWST} AGN in low-mass galaxies at $z > 4$.}
\label{histograms}
\end{figure}

Tens of overmassive BHs have been recently found by the \textit{JWST} at $z > 4$. We compile here all those AGN found by the \textit{JWST} based on the detection of broad H$\alpha$ or H$\beta$ emission in low-mass galaxies of log $M_\mathrm{*} \leq$ 10 M$_{\odot}$ at $z \sim 4-8$, shown as `JWST' in Figs.~\ref{MbhMstellar}-\ref{histograms}: \cite{2023ApJ...959...39H}, 10 sources; \cite{2023A&A...677A.145U}, one source; \cite{2023arXiv230801230M}, 12 sources; \cite{2023ApJ...953..180S}, one source; \cite{2024Natur.628...57F}, one source; \cite{2023ApJ...957L...7K}, one source; \cite{2023arXiv230904614Y}, two sources. In the case of H$\alpha$ emission (\citealt{2023ApJ...959...39H}; \citealt{2023A&A...677A.145U}; \citealt{2023arXiv230801230M}), the BH masses have been computed using the virial correlations of \cite{2013ApJ...775..116R} (see \citealt{2023ApJ...957L...3P}). For the \cite{2023ApJ...959...39H} sources, we have added in quadrature the typical 0.3 dex scatter of the virial relations to the instrumental BH mass error. In the case of H$\beta$ emission (\citealt{2023ApJ...953..180S}; \citealt{2024Natur.628...57F}; \citealt{2023ApJ...957L...7K}; \citealt{2023arXiv230904614Y}), the BH masses have been derived as for the VIPERS sources, using the relations from \cite{2006ApJ...641..689V} and \cite{2012ApJ...753..125S}, or from calibrations of \cite{2005ApJ...630..122G}. As for the VIPERS sources, the stellar masses of half of the \textit{JWST} sources (\citealt{2023A&A...677A.145U}; \citealt{2023arXiv230801230M}; \citealt{2023ApJ...957L...7K}) have been derived via SED fitting, allowing us only to distinguish between low-mass and massive galaxies. For the remaining \textit{JWST} sources (\citealt{2023ApJ...959...39H}; \citealt{2023ApJ...953..180S}; \citealt{2023arXiv230904614Y}), the stellar masses have been more reliably derived using spatial AGN-host decomposition. 

The total of 28 \textit{JWST} AGN in low-mass galaxies at $z > 4$ here compiled are offset from the local $M_\mathrm{BH}$-$M_\mathrm{*}$ correlation of active galaxies by $\Delta M_\mathrm{BH}$ = 2.8 $\pm$ 1.9 at a 5$\sigma$ level, confirming that they are overmassive. Both these \textit{JWST} sources and the VIPERS ones at $z = 0.35-2.7$ are indeed found to sit on the $M_\mathrm{BH}$-$M_\mathrm{*}$ at $z = 4-7$ derived by \citeauthor{2023ApJ...957L...3P} (2023, red line in Fig.~\ref{MbhMstellar}), which deviates at more than 3$\sigma$ confidence level from the local relation for AGN. We have run the same algorithm used to infer the $z > 4$ relation in \cite{2023ApJ...957L...3P}, adding the 19 low-mass galaxies investigated here at $z < 3$. We find the following values for the intercept $b$ and the slope $m$ of the linear relation: $b=-2.27 \pm 0.67$, and $m=1.10 \pm 0.07$. Remarkably, these values agree with the ones initially inferred in \cite{2023ApJ...957L...3P}: $b=-2.43 \pm 0.83$, and $m=1.06 \pm 0.09$. From a statistical perspective, this test suggests that the $z > 4$ sample discovered by \textit{JWST} and the VIPERS sample at $z < 3$ (including the new sources here reported and those from \citealt{2023ApJ...943L...5M}) belong to the same population.

The VIPERS sources at $z < 1$ of \cite{2023ApJ...943L...5M}, the new VIPERS sample at $z \sim 1-3$, and the \textit{JWST} sources at $z > 4$ also share the same range of $M_\mathrm{BH}$/$M_\mathrm{*}$ ratios (see Fig.~\ref{MbhMstellarevolution}), ranging from $\sim$0.2\% to $\sim$38\%. Most of the values are higher than the 1\% threshold used to define outlier sources (together with the $M_\mathrm{BH} > 10^7$ M$_{\odot}$ criterion, \citealt{2019MNRAS.485..396V}, which all of the VIPERS sources and most of the \textit{JWST} here compiled fullfill). Therefore, the AGN in low-mass galaxies at $z \sim 0.4-8$ considered here, both from the VIPERS and \textit{JWST} surveys, host BHs that are outliers with regard to the local $M_\mathrm{BH}$-$M_\mathrm{*}$ relation of active galaxies. This is independent of whether the $M_\mathrm{BH}$/$M_\mathrm{*}$ ratio evolves with redshift (\citealt{2010MNRAS.402.2453D}; \citealt{2010ApJ...708..137M}; \citealt{2010MNRAS.406L..35T}; \citealt{2011ApJ...742..107B}; \citealt{2020ApJ...888...37D}; \citealt{2018ApJ...867..148C}; \citealt{2024ApJ...964..154P}) or not (\citealt{2009ApJ...706L.215J}; \citealt{2011ApJ...741L..11C}; \citealt{2012ApJ...753L..30M}; \citealt{2015ApJ...802...14S}; \citealt{2021ApJ...909..188S}), see Fig.~\ref{MbhMstellarevolution}.

In addition to being overmassive, the \textit{JWST} AGN in low-mass galaxies at $z > 4$ also share a similar distribution of bolometric luminosities as the VIPERS sources at $z < 1$ and the new VIPERS sample at $z \sim 1-3$ (see Fig.~\ref{histograms}, top panel), with log L$_\mathrm{bol}$ = 43.7 to 47.2 erg s$^{-1}$, and similar distribution of Eddington ratios (see Fig.~\ref{histograms}, bottom panel), with most of the sources accreting at sub-Eddington rates (ie., $\lambda_\mathrm{Edd}<$ 1). Applying a Mann-Whitney statistical test between the VIPERS sample (at $z \sim 0.4-3$) and the \textit{JWST} sample (at $z > 4$) returns a p-value greater than 0.02 both for the Eddington ratio and the bolometric luminosity distributions, hence we cannot reject the null hypothesis that the two samples are drawn from the same distribution at a 98\% confidence level. We note though that performing a robust statistical test to compare two independent samples when the sample size is very small is challenging. Yet, the statistical results seem to be consistent with the visual inspection of the histograms presented in Fig.~\ref{histograms}, where we see that the central tendencies, spread, and shapes between the samples are similar.
All these low-mass galaxies hosting overmassive AGN could thus be similar sources simply detected at different cosmic epochs.

\section{Discussion and Conclusions}
\label{conclusions}
In this Letter, we report on the first sample of significantly overmassive SMBHs detected at cosmic noon, i.e., in the redshift range $z=1-3$. Remarkably, these overmassive systems share several similar physical properties (BH mass, stellar mass, bolometric luminosity, Eddington ratio) with their high-$z$ counterparts, found ubiquitously in recent \textit{JWST} surveys. 

\cite{2024arXiv240307986E} recently investigated four \textit{JWST} fields at $z >$ 6 and found that the duty cycle for UV-luminous quasars at these cosmic epochs is significantly less than unity. This suggests that such high-z SMBHs may undergo episodic and highly dust-obscured phases of radiatively inefficient super-Eddington accretion. In our `cosmic noon' sample of AGN low-mass galaxies as well as that at $z < 1$, all the sources are characterized by sub-Eddington luminosities; most have an Eddington ratio lower than 60\% Eddington. Our sources are all observed at $z < 3$ and are intrinsically less likely to undergo extreme super-Eddington phases, as the availability of cold gas necessary to fuel such extreme growth phases is much lower at these redshifts than at $z > 6$ (see, e.g., \citealt{2010MNRAS.406...43P}). Furthermore, we point out that \cite{2024arXiv240303872J} recently presented the discovery at $z >$ 6 of an extremely overmassive BH, with a mass of $\sim 40\%$ the total stellar mass of the host galaxy, which is accreting at the meager rate of $2\%$ Eddington. Hence, extraordinarily overmassive yet low-luminosity SMBHs have already been detected in the early Universe. 

In a recent study of variability-selected AGN in low-mass galaxies at $z\sim 0.5-4$ drawn from the COSMOS survey (\citealt{2007ApJS..172....1S}) and followed up in the Hyper Suprime-Cam Subaru Strategic Program (HSC-SSP; \citealt{2022PASJ...74..247A}) - a sample coincident in redshift range with the `cosmic noon' sample studied here - \cite{2024arXiv240206882B} find that these sources are also overmassive compared to the local $M_{\rm BH}-M_*$ relation for active galaxies. However, they appear to be consistent with the local relation for inactive early-type galaxies, as would be the case for some of the  overmassive BHs here presented (at cosmic noon and at $z>4$ with \textit{JWST}). 
Local inactive galaxies with dynamical BH mass measurements tend however to be biased by angular resolution-related selection effects (e.g., \citealt{2007ApJ...660..267B}; \citealt{2016MNRAS.460.3119S}). Using the resolution-corrected or debiased correlation of local quiescent galaxies of \citeauthor{2016MNRAS.460.3119S} (2016; see also \citealt{2019MNRAS.485.1278S}, eq. 5) we find that both the AGN in low-mass galaxies at cosmic noon and at $z < 1$ as well as the \textit{JWST} ones at $z > 4$ are clearly offset from the correlation (see Fig.~\ref{MbhMstellar}). Such offset seems to be softened for the \textit{JWST} sources at $z > 4$ when using the $M_{\rm BH}-\sigma$ correlation, where $\sigma$ is the stellar velocity dispersion derived from the width of the [OIII] emission line (e.g., \citealt{2024arXiv240303872J}). The absence of such line in the VIPERS spectra of the $z \sim 1-3$ sources prevents probing this for the `cosmic noon' sample. 

With $\sim 3$ Gyr of cosmic time elapsed between the median redshift of our `cosmic noon' sample and the \textit{JWST}'s high-$z$ sample, it is challenging to connect these samples in a straightforward fashion as the causes for departure from the local scaling relation are likely to be different at these cosmic epochs. Nonetheless, some considerations are worth mentioning.

At high redshifts, the existence of overmassive SMBHs can be attributed to seeding physics. 
Recent cosmological simulations (e.g., \citealt{2023MNRAS.519.2155S}) suggest that heavy seeding, for example as a result of the direct collapse BH seed formation scenario, will result in an early (i.e., $z > 15-20$) ratio of BH to stellar mass close to unity (\citealt{2013MNRAS.432.3438A}). In other works (e.g., \citealt{2024arXiv240203626B}) lower mass seeds of $\leq 10^3$ M$_{\odot}$ are already able to predict BH masses $\sim$10-100 times higher than expected from local scaling relations. By cosmic noon, however, we expect two additional important effects to modulate the scaling relations between central SMBHs and the stellar content of their host galaxies: (i) feedback processes and (iii) impact of interactions/mergers leading to tidal stripping. 

First, we note an important caveat: selection effects might well be operating here. For instance, we do know that the highest redshift ($z>6$) \textit{JWST} detected sources are likely the most luminous population and are therefore outliers in terms of their luminosities and inferred BH masses. It is conceivable that these extremely bright sources are preferentially overmassive. It has been argued persuasively at least in the case of the $z=10.1$ source UHZ1 that this system was likely seeded with a massive seed of $10^4-10^5$ M$_{\odot}$ causing it to be overmassive (\citealt{2024NatAs...8..126B}; \citealt{2024ApJ...960L...1N}). So heavy seeding coupled with feedback physics as noted by \cite{2024ApJ...964..154P} might be implicated for these extremely high-redshift sources. However, the fact that we still see overmassive systems $\sim 3$ Gyr after in cosmic time could suggest that there exists a galaxy population that even by these late times was not able to build up enough stellar mass to shift towards the local $M_\mathrm{BH} - M_\star$ relation. 

The finding that these systems are still overmassive at cosmic noon suggests that BH feedback processes are playing a significant role in shaping their host galaxies (\citealt{2022MNRAS.516.2112K}), likely quenching star formation and stellar growth (e.g, \citealt{2024ApJ...964..154P}; \citealt{2024ApJ...961L..39S}). This could explain the compactness and dust-reddened emission of most of the high-$z$ \textit{JWST} sources (\citealt{2024ApJ...964...39G}). Indeed, the presence of AGN outflows is prominent in one of the most distant AGN (GN-z11; \citealt{2024Natur.627...59M}) and in some of the \cite{2023ApJ...959...39H} sources here considered (for which \citealt{2023ApJ...959...39H} correct the H$\alpha$ line emission used to derive the BH mass). 

In the `cosmic noon' sample here reported, five of the sources show possible asymmetric components in the Ly$\alpha$, CIV or CIII] emission indicative of outflows. We note that in the local Universe evidence is growing that AGN feedback can be equally or even more important than supernova feedback in shaping dwarf galaxies (\citealt{2019ApJ...884...54M}; \citealt{2019MNRAS.488..685M}; \citealt{2020ApJ...905..166L}; \citealt{2022MNRAS.511.4109D}; \citealt{2022Natur.601..329S}), and AGN feedback is expected to have impacted BH growth in dwarf galaxies across cosmic time (\citealt{2019NatAs...3....6M}). 

A more comprehensive study of the detailed physics of AGN feedback processes in the overmassive systems at cosmic noon and beyond is required to better understand the co-evolution and the mass assembly history of stars and the SMBHs hosted in galaxies. In particular, the role of outflows in modulating asynchronous BH-galaxy growth and/or growth in tandem is urgently needed. In addition, as noted in recent analysis of simulations (e.g., \citealt{2023MNRAS.522.4963W}), environment is also likely to play an important role in determining where accreting sources fall in relation to the local $M_{\rm BH}-M_*$ relation. A recent study by \cite{2024arXiv240214706I} suggests that dust-rich environments in the high-redshift Universe could create conditions prone to generate intrinsically overmassive populations of BHs, with a distribution similar to what our data at $z < 3$ is showing. Future follow-up observational studies to characterize the immediate environments for signs of mergers and interactions would enable us to derive a full picture of the many competing effects that operate in co-evolution.

\section*{Acknowledgments}
The authors thank the anonymous referee for insightful comments. The authors thank Richard Mushotzky for providing important information regarding the X-ray detection of some of the sources.
The authors would like to express their appreciation to Jen Miller at Gemini Observatory for providing reduced GNIRS data for one of our targets.  
M.M. acknowledges support from the Spanish Ministry of Science and Innovation through the project PID2021-124243NB-C22. This work was partially supported by the program Unidad de Excelencia Mar\'ia de Maeztu CEX2020-001058-M.
F.P. acknowledges support from a Clay Fellowship administered by the Smithsonian Astrophysical Observatory. This work was also supported by the Black Hole Initiative at Harvard University, which is funded by grants from the John Templeton Foundation and the Gordon and Betty Moore Foundation. 
HS is supported by the international Gemini Observatory, a program of NSF’s NOIRLab, which is managed by the Association of Universities for Research in Astronomy (AURA) under a cooperative agreement with the National Science Foundation, on behalf of the Gemini partnership of Argentina, Brazil, Canada, Chile, the Republic of Korea, and the United States of America. PN acknowledges support from the Gordon and Betty Moore Foundation and the John Templeton Foundation that fund the Black Hole Initiative (BHI) at Harvard University where she serves as one of the PIs.
This work has been supported by the Polish National Agency for Academic Exchange (Bekker grant BPN/BEK/2021/1/00298/DEC/1), the European Union's Horizon 2020 Research and Innovation programme under the Maria Sklodowska-Curie grant agreement (No. 754510).

Based on observations obtained at the international Gemini Observatory, a program of NSF’s NOIRLab, GN-2021A-FT-216, which is managed by the Association of Universities for Research in Astronomy (AURA) under a cooperative agreement with the National Science Foundation on behalf of the Gemini Observatory partnership: the National Science Foundation (United States), National Research Council (Canada), Agencia Nacional de Investigaci\'{o}n y Desarrollo (Chile), Ministerio de Ciencia, Tecnolog\'{i}a e Innovaci\'{o}n (Argentina), Minist\'{e}rio da Ci\^{e}ncia, Tecnologia, Inova\c{c}\~{o}es e Comunica\c{c}\~{o}es (Brazil), and Korea Astronomy and Space Science Institute (Republic of Korea).
This work was enabled by observations made from the Gemini North telescope, located within the Maunakea Science Reserve and adjacent to the summit of Maunakea. We are grateful for the privilege of observing the Universe from a place that is unique in both its astronomical quality and its cultural significance.

\bibliographystyle{aasjournal}
\bibliography{HighzAGN_astroph.bib}

\clearpage
\appendix

\begin{figure}
\centering
\includegraphics[width=0.99\textwidth]{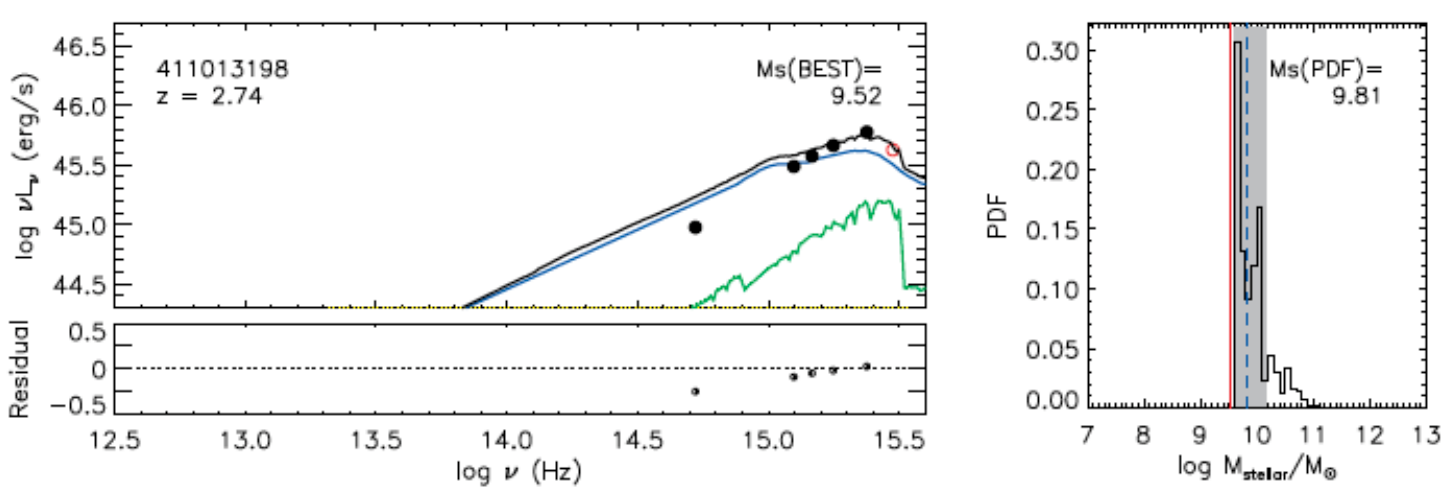}
\includegraphics[width=0.99\textwidth]{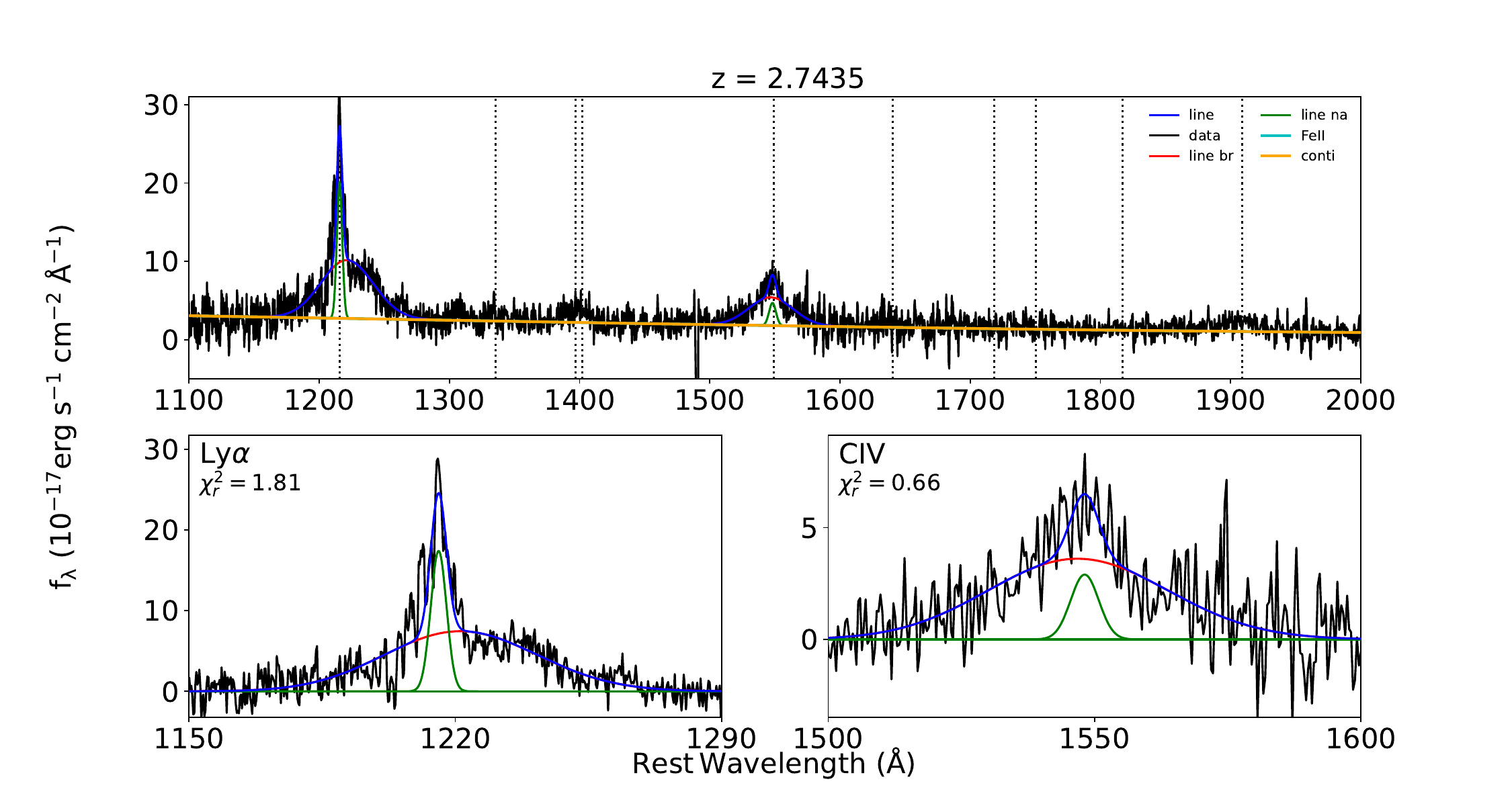}
\includegraphics[width=0.3\textwidth]{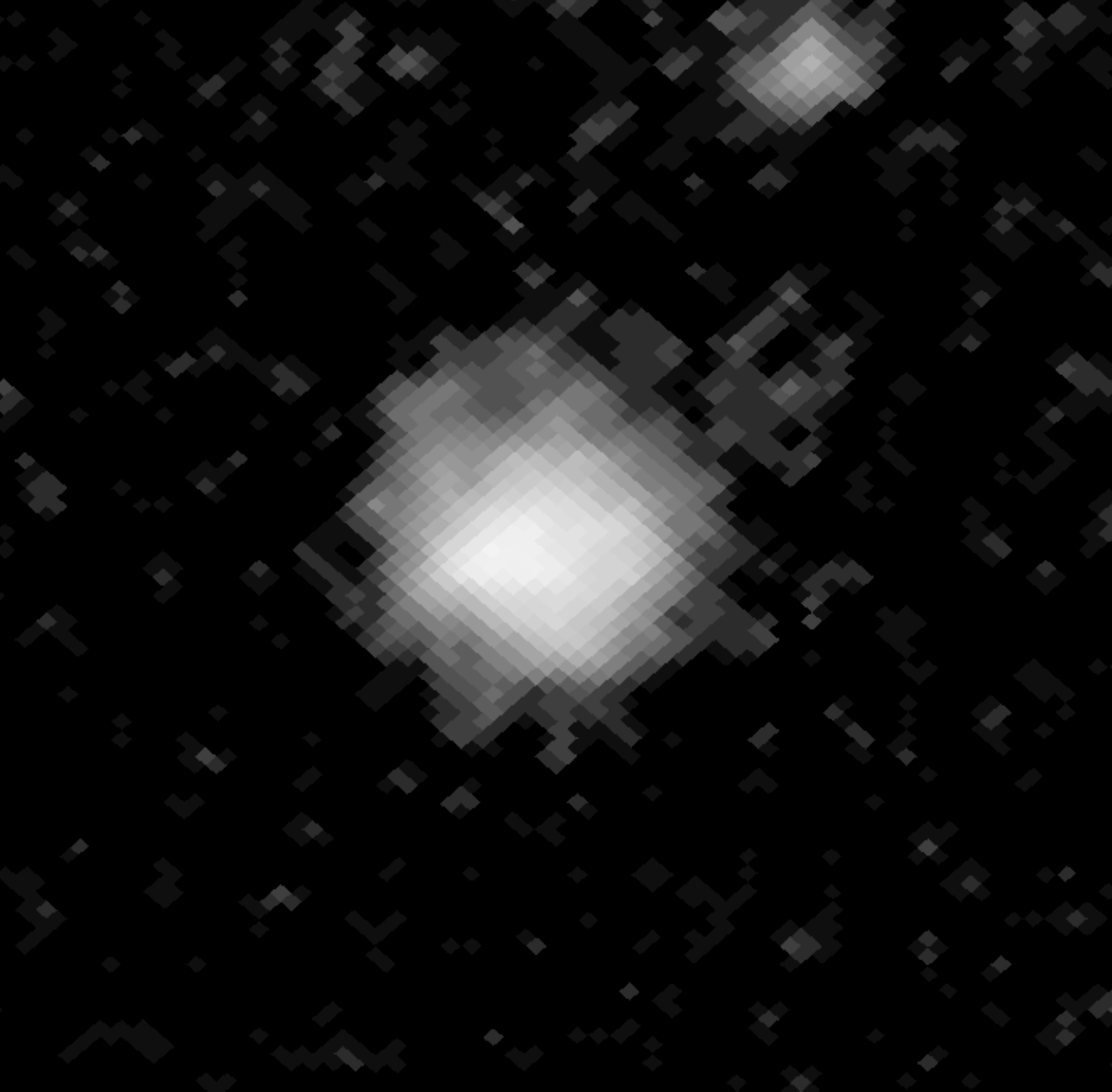}
\caption{One of the 12 `cosmic noon' low-mass galaxies hosting an overmassive BH. Top left: Spectral energy distribution fitting of the rest-frame observed ultraviolet, optical, and infrared (when available) photometry (black points) with the best-fit model (black curve), including a combination of the galaxy template (green), an AGN accretion disk component (blue) and an AGN dust torus model (yellow). Top right: PDF for the stellar mass taking into account all possible fractions of AGN emission and providing an upper limit on the stellar mass. So the most probable value (MsPDF, blue dashed line) has a higher value than the best-fit stellar mass (MsBEST, red solid line). The 16 and 84 percentile intervals (gray shades) are also indicated. Middle: Emission line fitting of the VIPERS spectrum including the continuum emission (in yellow, top panel), and the broad lines (in blue) decomposed into broad (in red) and narrow (in green) components (zoom-in in the bottom panels). Bottom: Subaru Hyper Suprime-Cam image in the i-band. }
\label{SEDfit}
\end{figure}

\section{Gemini observations}
\label{Gemini}
We obtained a 0.8-2.5$\mu$m spectrum of 401126746 using the Gemini Near-Infrared Spectrograph (GNIRS) on the Gemini telescope with queue-observing mode, under the program GN-2021A-FT-216. GNIRS was configured using a Short Blue camera (0,15"/pixel), 32 l/mm cross-dispersed mode, and a 0.675" wide slit. Observations utilized the standard ABBA method of nodding along the slit to enable sky subtraction. We also observed telluric standard stars before and after the observations.
The data were reduced using the XDGNIRS pipeline developed by the Gemini Observatory, based on the Gemini IRAF package. The reduction pipeline includes standard image cleaning for pattern noise and artifacts, flatfielding, sky subtraction, distortion correction, and rectification of 2D data. The arc spectra were used for wavelength calibration. The spectrum of the telluric standards was processed in a similar way, followed by the removal of intrinsic hydrogen absorption lines, and used for the telluric corrections and flux calibration. 

\end{document}